\documentclass{PoS}

\usepackage{amsmath}
\usepackage{amssymb}
\usepackage{cite}
\usepackage{dsfont}
\usepackage{epsfig}
\usepackage{latexsym}
\usepackage{booktabs}
\usepackage{graphicx}% Include figure files
\usepackage{feynmp-auto}

\newcommand{\be}{\begin{equation}}
\newcommand{\ee}{\end{equation}}

\newcommand{\ba}{\begin{eqnarray}}
\newcommand{\ea}{\end{eqnarray}}
\renewcommand{\eqref}[1]{Eq.~(\ref{#1})}

\newcommand{\figref}[1]{Fig.~\ref{#1}}

\title{A variational method for spectral functions}

\ShortTitle{A variational method for spectral functions}

\author{Tim Harris~\textsuperscript{$a$}, Harvey B. Meyer~\textsuperscript{$a$}, \speaker{Daniel Robaina}~\textsuperscript{$a,b$} \\

\textsuperscript{a}PRISMA Cluster of Excellence,
Institut f\"ur Kernphysik and Helmholtz~Institut~Mainz,
Johannes~Gutenberg-Universit\"at~Mainz,
D-55099 Mainz, Germany.\\

\textsuperscript{b}Institut f\"ur Kernphysik,
Technische Universit\"at Darmstadt,\\ 
Schlossgartenstrasse 2, D-64289 Darmstadt, Germany.\\
E-mail: \email{robaina@theorie.ikp.physik.tu-darmstadt.de}
}
        
\abstract{The Generalized Eigenvalue Problem (GEVP) has been used extensively in the past in order to reliably extract energy levels from time-dependent Euclidean correlators calculated in Lattice QCD. We propose a formulation of the GEVP in frequency space. Our approach consists of applying the model-independent Backus-Gilbert method to a set of Euclidean two-point functions with common quantum numbers. A GEVP analysis in frequency space is then applied to a matrix of estimators that allows us, among other things, to obtain particular linear combinations of the initial set of operators that optimally overlap to different local regions in frequency. We apply this method to lattice data from NRQCD. This approach can be interesting both for vacuum physics as well as for finite-temperature problems.}

\FullConference{34th annual International Symposium on Lattice Field Theory\\
		 24-30 July 2016\\
		 University of Southampton, UK}

\begin{document}
\unitlength = 1mm

\section{Introduction}
Extracting energy-levels from Euclidean correlators in a finite four dimensional box is a problem for which the Lattice community has devoted significant amount of computer and human resources. This has become specially relevant due to the work of M.~L\"uscher \cite{luescher} relating the discrete energy spectrum of a given theory in finite volume to scattering amplitudes of the same theory in the infinite volume limit. In order to extract different phase shifts, one needs on the lattice side a precise and reliable procedure to read off the energy-levels from effective mass plateaus. The most widely used method is called the Generalized Eigenvalue Problem (GEVP) \cite{gevp}. The main idea consists in building a basis of operators with common quantum numbers that consequently interpolate the same discrete but infinite tower of energy-eigenstates. From the additional information of the off-diagonal elements of the matrix of correlators (let us call it $G_{\alpha \beta}(\tau)$ where $\alpha,\beta=1,\dots,N$ with $N$ the number of basis operators included and $\tau$ is the Euclidean time coordinate) one can construct particular linear combinations that optimally couple to the ground state, first excited state, etc. This reduces the contribution of higher states providing effectively longer and more trustworthy plateaus with consequently smaller errors on the final results.

This set of ideas and methods, which have been used extensively in the study of vacuum properties of QCD, should ideally be carried over to situations where isolating individual energy eigenstates is not feasible, either because the spectrum is too dense or because one is considering QCD at finite temperature. This is the aim of this very preliminary study. When considering QCD in the presence of a thermal bath the relevant degrees of freedom (quasiparticles) have modified properties compared to the vacuum situation due to medium interactions (see as an example our recent study about the pion quasiparticle below the phase transition \cite{rob1,rob2}). These modifications can be understood in terms of changes in the spectral functions. Moreover, in the plasma phase, where hadronic bound states dissociate, we expect the collective behavior of quarks and gluons to be responsible for the non-equilibrium properties, such as the emission rate of soft photons or the dilepton production rate \cite{cms}. Via Kubo formulae, the transport coefficients can be related to the small-frequency domain of the thermal spectral functions.

Therefore, extracting spectral functions from Euclidean QCD is one of the most important goals of the finite temperature lattice community. Euclidean correlators $G_E(\tau)$ and their corresponding spectral functions $\rho(\omega)$ are related via the following integral equation
\be
G_E(\tau) = \int d\omega \rho(\omega) \underbrace{\frac{\cosh(\omega(\beta/2 - \tau))}{\sinh(\omega \beta/2)}}_{K(\omega,\tau)}
\label{eq:main}
\ee
where $K(\omega,\tau) \to e^{-\omega \tau}$ when $1/T = \beta \to \infty$. Unfortunately, the inversion of this equation is numerically an ill-defined problem given the limited amount of data available from typical lattice calculations (see e.g. \cite{harvey} Sec. 5). It is necessary to find alternative methods to distinguish the gross features of the spectral function, such as resonance peaks or smooth backgrounds. We will exploit the Backus-Gilbert-method (BGM) \cite{bgm}, recently applied to finite-temperature QCD in \cite{rob2,bgm1,bgm2} because we have to use a linear method in conjunction with the GEVP.

\section{GEVP for spectral functions}
\subsection{Backus-Gilbert method (BGM)}
The goal is to invert \eqref{eq:main} for a given kernel $K(\omega,\tau)$ and input data $G(\tau_i)$ (we drop the subscript `$E$' from now on). The main idea of the BGM is to define an estimator $\hat{\rho}(\bar{\omega})$
\be
\hat{\rho}(\bar{\omega}) = \int^\infty_0 d\omega \hat{\delta}(\bar{\omega},\omega) \rho(\omega)
\label{eq:hatrho}
\ee
where $\hat{\delta}(\bar{\omega},\omega)$ is called the resolution function. It is a smooth function peaked around $\bar{\omega}$, normalized according to $\int^\infty_0 d\omega \hat{\delta}(\bar{\omega},\omega) = 1$, and parametrized by a priori unknown coefficients $q_i(\bar{\omega})$
\be
\hat{\delta}(\bar{\omega},\omega) = \sum^{N_T}_{i=1} q_i(\bar{\omega}) K(\omega, \tau_i).
\ee
Minimizing the second moment of $[\hat{\delta}(\bar{\omega},\omega)]^2$, equivalent to maximizing the resolution for a fixed value of $\bar{\omega}$, subject to the condition that the area is 1, completely fixes the coefficients $q_i(\bar{\omega})$ at every $\bar{\omega}$ (details can be found in \cite{rob2} Sec. E and \cite{nr}). Fig.\ 1 displays examples of resolution functions centered at different values of $\bar{\omega}$ for the data discussed in the following sections. In contrast to other methods like the popular Maximum Entropy Method (MEM), the BGM is a linear method, meaning that the estimator for the spectral function is a linear combination of the points of the Euclidean correlator
\be
\hat{\rho}(\bar{\omega}) = \sum^{N_T}_{i=1} q_i(\bar{\omega}) G(\tau_i).
\ee
In practice, the resolution is only limited by the number and quality of the input data and consequently no prior ansatz is needed for $\hat{\rho}(\bar{\omega})$ which is a smoothened version of the true spectral function. In the following we concentrate on combining the BGM with the GEVP. For an extensive analysis of the BGM on mock-data see \cite{rob2} Appendix C.
\begin{center}
\begin{figure}
\centering
\includegraphics[width=.5\textwidth]{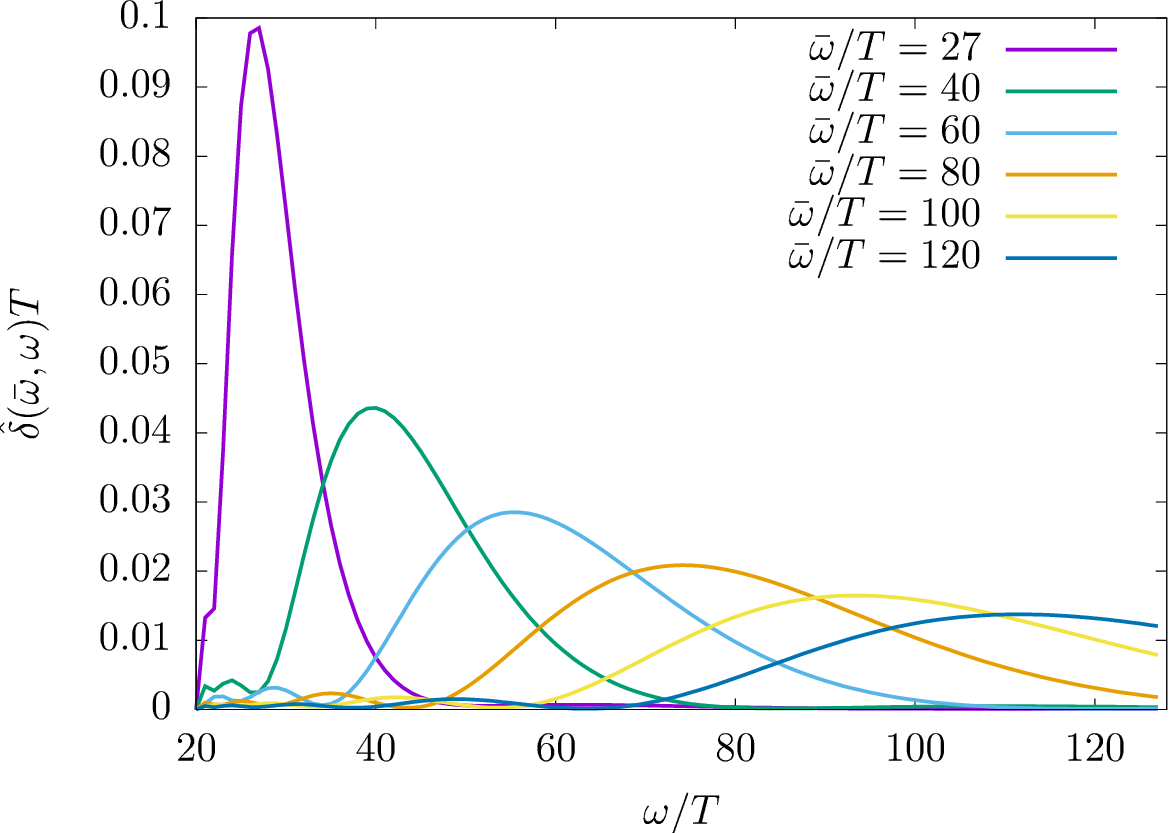}
\caption{Resolution functions $\hat{\delta}(\bar{\omega},\omega)$ for different choices of $\bar{\omega}$ The kernel is $K(\omega, \tau)=e^{-\omega t}\sqrt{\omega}$. The resolution is greater for smaller $\bar{\omega}$, as expected.}
\label{fig:res}
\end{figure}
\end{center}

\subsection{Frequency space GEVP}
Suppose that we now construct a matrix of euclidean correlators 
\be
G_{\alpha \beta}(\tau_i) = \int d^3x \left<O_\alpha(\tau_i,\vec{x}) O^\dagger_\beta(0)\right>, \qquad \alpha,\beta=1,\dots,N.
\ee
We then apply the BGM method on each entry of the matrix with a common resolution function $\hat{\delta}(\bar{\omega},\omega)$ to obtain $\hat{\rho}_{\alpha \beta}(\bar{\omega})$. We solve for the eigenvalues and eigenvectors of the following generalized eigenvalue equation
\be
\hat{\rho}_{\alpha \beta}(\bar{\omega})v^{(i)}_\beta(\bar{\omega}) = \lambda^{(i)}(\bar{\omega})N_{\alpha \beta}v^{(i)}_\beta(\bar{\omega}) \qquad i=1,\dots,N.
\label{eq:gevp}
\ee
We sort the eigenvalues in ascending order. The metric $N_{\alpha \beta}$ must transform under a linear change of the operator basis $P$ like $P^\dagger N P$ and in the following we choose $N_{\alpha \beta} = G_{\alpha \beta} (\tau_0)$ analogously to the coordinate space GEVP. In the following we test the method on real lattice zero temperature NRQCD data in the $\Upsilon$-channel.

\section{Example: $\Upsilon$-meson from NRQCD}
NRQCD provides a good testing ground for spectral reconstruction because the data is precise and the kernel $K(\omega,\tau)$ takes the form $e^{-\omega \tau}$ for any temperature. We consider the $\Upsilon$-channel with a basis of operators of dimension 3 $(N=3)$. The lattice parameters are shown in Table 1 and the operators used are written in Table 2. From pertubative results it is known that these types of spectral functions grow as $\omega^{1/2}$ in the UV \cite{burnier}. It is therefore common practice when using the BGM to reweight with the asymptotic behavior due to the fact that the method is exact if the target spectral function is constant. In this specific setup \eqref{eq:main} and \eqref{eq:hatrho} read respectively 
\begin{align}
G_{\alpha \beta}(\tau_i) &= \int^\infty_0 \left(\frac{\rho_{\alpha \beta}(\omega)}{\sqrt{\omega}}\right)\underbrace{\left( e^{-\omega \tau_i} \sqrt{\omega}\right)}_{K(\omega,\tau_i)},\\
\hat{\rho}_{\alpha \beta}(\bar{\omega}) &= \int^\infty_0 d\omega \hat{\delta}(\bar{\omega},\omega)\left( \frac{\rho_{\alpha \beta}(\omega)}{\sqrt{\omega}}\right).
\end{align}
\begin{table}
\begin{center}
\begin{tabular}{cccccc}
$N_f$ & $\beta/a$ & $a_s$ & $1/a_\tau$ & $m_\pi/m_\rho$ & $m_\pi L$ \\ \hline
$2+1$ & $128$ & $0.12$fm & $5.67$GeV & $0.45$ & $3.9$ \\ \hline 
\end{tabular}
\label{tab:1}
\caption{Ensemble parameters for anisotropic lattices generated by the HadSpec Collaboration \cite{hadspec1,hadspec2}.}
\end{center}
\end{table}
\begin{table}
\begin{center}
\begin{tabular}{ccr}
$\alpha$ & $\psi_\alpha(x)$ &source type \\ \hline
1 & $\psi(x)$ & point\\
2 & $\sum_{\vec{y}} e^{-(\vec{x}-\vec{y})^2/\sigma^2} \psi(\vec{y})$ & gaussian smeared\\
3 & $\sum_{\vec{y}}\left(4 \frac{(\vec{x}-\vec{y})^2}{\sigma^2}-3\right) e^{-(\vec{x}-\vec{y})^2/\sigma^2} \psi(\vec{y})$ & excited smeared\\ \hline
\end{tabular}
\label{tab:2}
\caption{Basis of NRQCD interpolators. The operators are constructed as $O^\dagger_\alpha (x) = \chi^\dagger_\alpha(x) \sigma_1 \psi_\alpha(x)$.}
\end{center}
\end{table}
\figref{fig:res_rho} shows the diagonal elements of $\hat{\rho}_{\alpha \beta} (\bar{\omega})$ together with the $\bar{\omega}$-behavior of the three eigenvalues (bottom-right panel). As expected, the point source couples to the whole spectrum, while the gaussian smeared suppresses excited states contamination and enhances the spectral weight of the ground state. The "excited smeared" source clearly reveals a first excited state. The peak positions coincide with the energies from a multi-exponential fit to the correlation functions. 

\section{Discussion and Conclusions}
\begin{center}
\begin{figure}
\centering
\includegraphics[width=.47\textwidth]{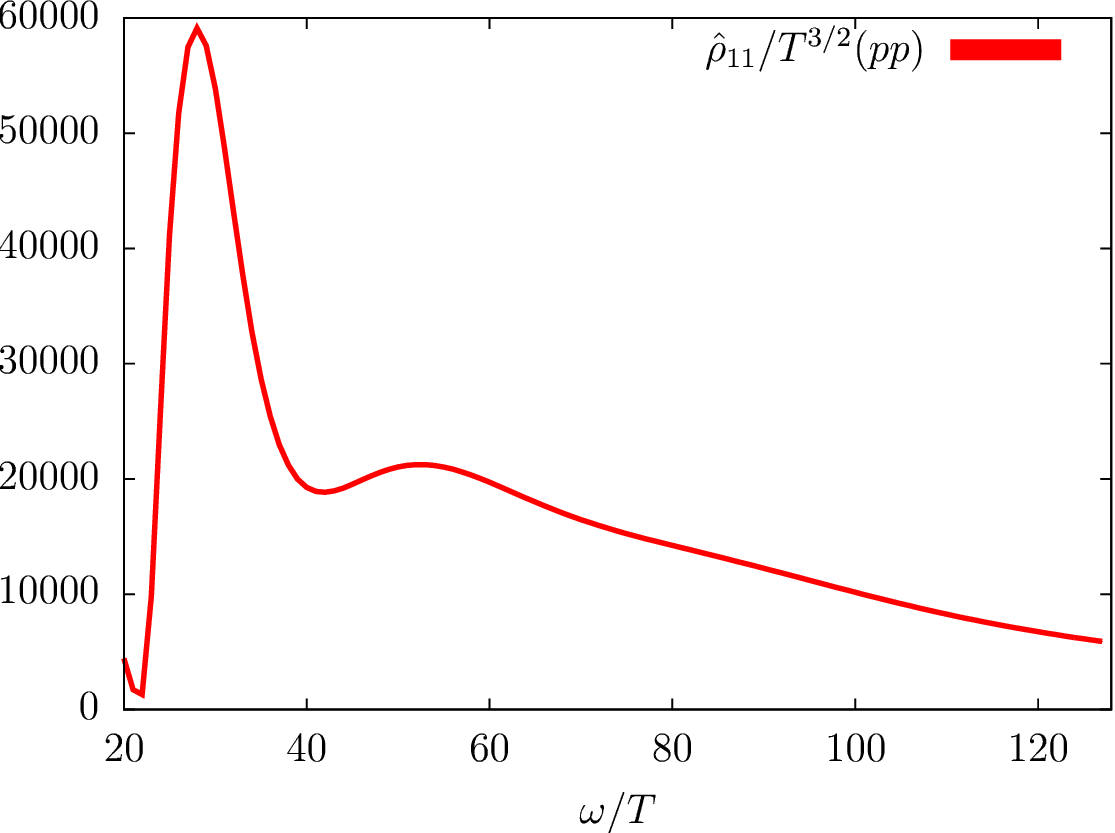}
\hspace{.7cm}
\includegraphics[width=.47\textwidth]{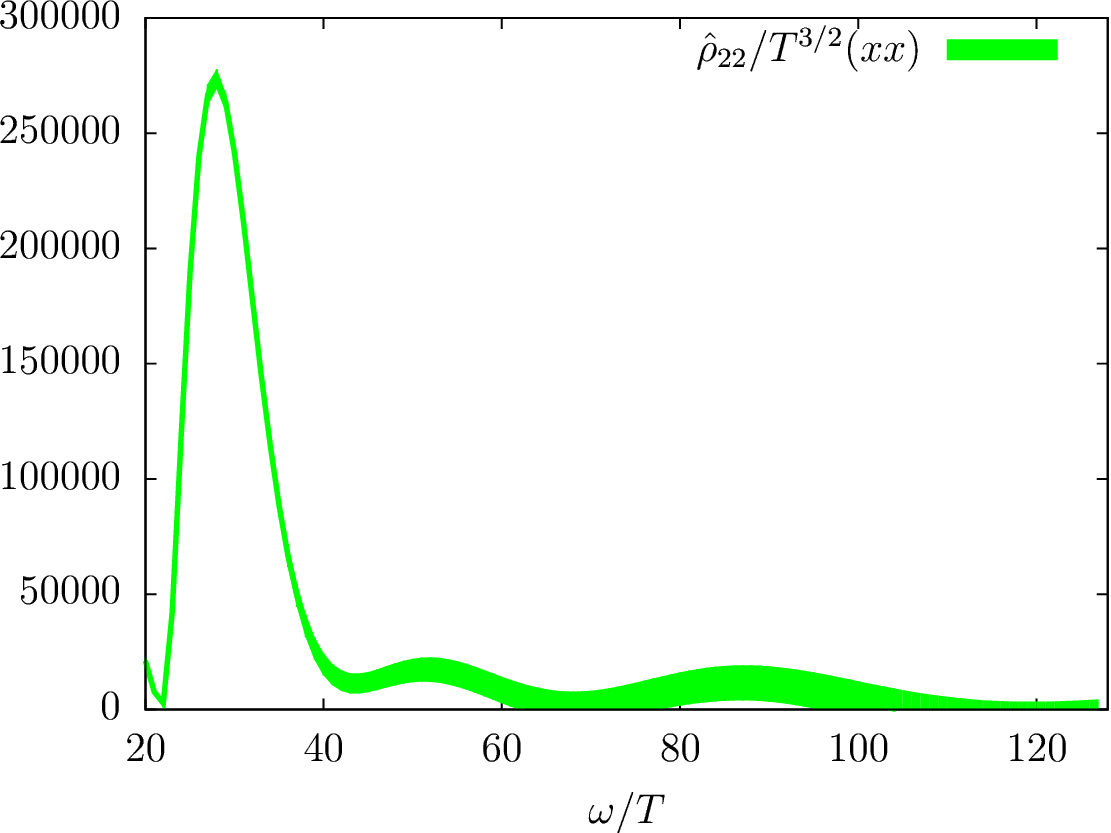}
\includegraphics[width=.47\textwidth]{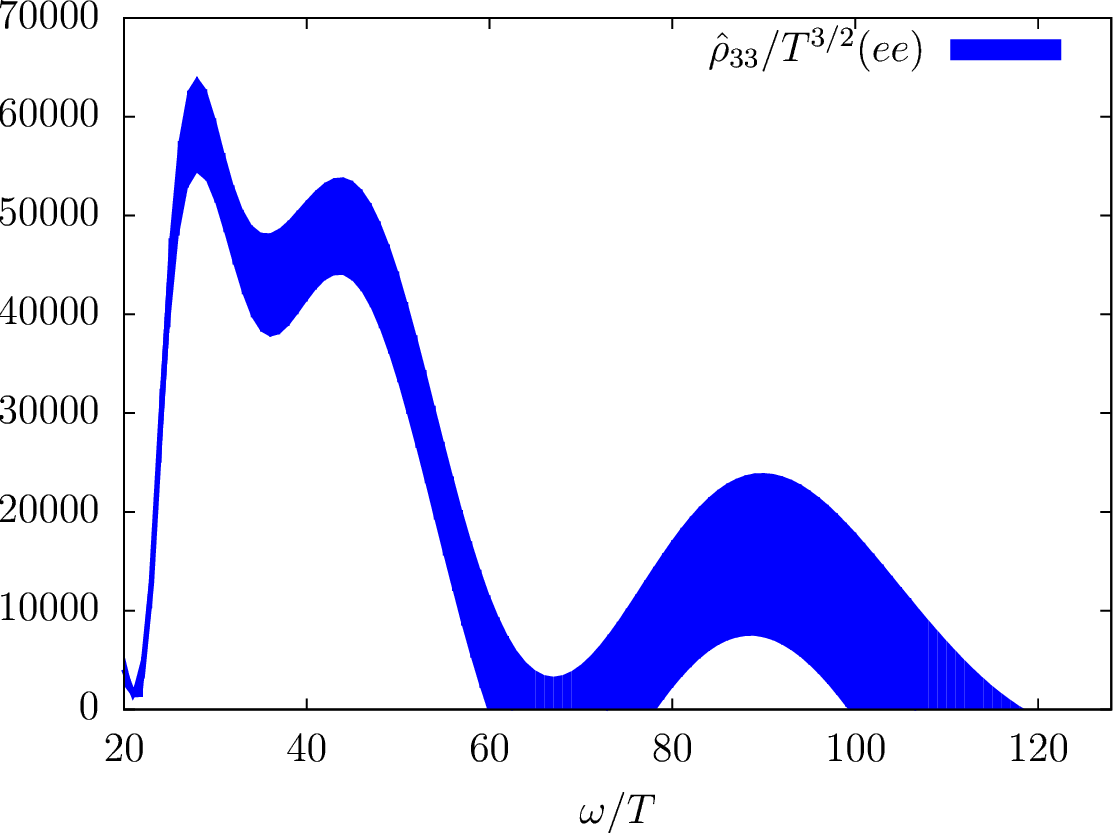}
\hspace{.7cm}
\includegraphics[width=.47\textwidth]{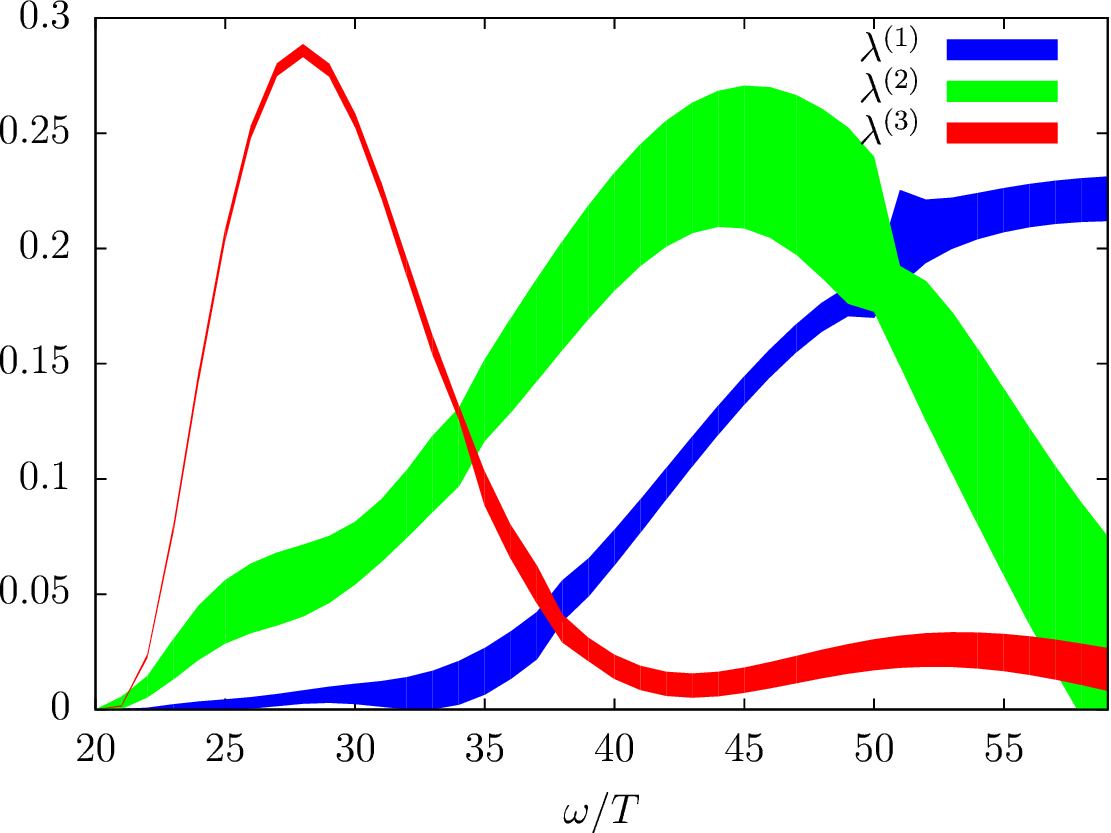}
\caption{Diagonal elements of $\hat{\rho}_{\alpha \beta} (\bar{\omega})$. Bottom-right: Eigenvalues calculated through GEVP solving of \protect\eqref{eq:gevp}.}
\label{fig:res_rho}
\end{figure}
\end{center}

As explained in previous sections, the BGM resolves the spectral function locally in frequency space. It yields the estimator $\hat{\rho}(\bar{\omega})$ by recomputing the coefficients $q_i$ at every $\bar{\omega}$. Therefore, analogously to the GEVP in coordinate space, it may be possible to construct optimal operators
\be
O^{\text{opt}}(\bar{\omega}^{(N)},x) = v^{(N)}_\alpha(\bar{\omega}) O_\alpha(x)
\ee
that strongly overlap onto specific regions in frequency space and then study the cross correlator of the optimized operator with a local operator such as the conserved vector current. In particular, this correlator may help determine the coupling of a photon to a resonance or thermal quasiparticle. Moreover, recalling the formula for the spectral decomposition
\be
G_{\alpha \beta}(\tau) = \sum_{n} Z^{(n)}_{\alpha \beta} e^{-E_n \tau}, \qquad Z^{(n)}_{\alpha \beta} = Z^{(n)}_\alpha Z^{(n)*}_\beta, \qquad Z^{(n)}_\alpha = \left<0|O_\alpha(0)|n\right>
\ee
one sees that the matrix $Z^{(n)}$ is hermitian and has $\text{rank}(Z)=1$. In an idealized situation we therefore expect the eigenvalues coming from \eqref{eq:gevp} to show an approximate rank-one behavior (i.e.\ $\lambda^{(N)}\gg \lambda^{(N-1)}$) whenever $\bar{\omega}$ coincides approximately with some energy-level $E_n$ provided that the width of the resolution function is small enough that only one state is resolved at the time. Indeed, looking at the bottom-right panel of \figref{fig:res_rho} we observe that at the $\omega$-values corresponding to the ground and first excited states, the eigenvalues show a hierarchy in its magnitude. We suggest that this criterion can be used to identify quasiparticles.

In the future, one might want to investigate how the number of operators affects our results. We are currently working on bigger bases ($N\simeq 10$) with QCD ensembles at zero and non-zero temperature. Eventually, we want to study our method perturbatively. In order to have a more rigorous and phenomenologically realistic testing ground to perform numerical mock-data analyses we provide in the Appendix a simple 1-loop QFT continuum, infinite volume calculation of a spectral function containing a resonance. In addition we want to investigate how the finite volume spectral function (a sum of delta-functions) tends to a continuous form in the thermodynamic limit (see \cite{harvey} Sec. 3). The BGM seems ideal for this purpose since only after a convolution of the finite volume spectral function with a resolution function, the limit of $\rho(\omega)$ when $L\to \infty$ is well-defined.     

\appendix
\section{Perturbative spectral function calculation from a resonance model}
We consider the Lagrangian density of two dynamical scalar fields $K$ and $\phi$ in four dimensions and Euclidean metric with different masses $m_K$ and $m_\phi$. We allow for all dimension four operators that respect a $Z_2$-symmetry $\phi \mapsto -\phi$ which makes the theory renormalizable. We demand also $m_K > 2 m_\phi$ such that the $K$-particle can decay into two $\phi$-particles. The Lagrangian reads
\begin{align}
\mathcal{L} &= \frac{1}{2} (\partial \phi)^2 + \frac{1}{2}m_\phi^2 \phi^2 + \frac{1}{2} (\partial K)^2 + \frac{1}{2}m_K^2 K^2 + \mathcal{L}_\text{int} \nonumber \\
\mathcal{L}_{int} &= \frac{1}{4!} \lambda \phi^4 + \frac{1}{2}g\phi^2 K + \frac{1}{4!}\lambda_k K^4 + \frac{1}{3!} g_k K^3 + \frac{1}{4}h K^2 \phi^2
\end{align}

As a first exercise we study the $KK$-Euclidean correlation function in momentum space defined as
\be
\int d^4x \, e^{ipx} \left<K(x)K(0)\right> = \frac{1}{p^2 + m^2_K - \Sigma_K(p^2)}.
\ee
Our on-shell renormalization scheme demands $\text{Re}\Sigma_K(p^2=-m^2_K) = \text{Re}\Sigma'_K(p^2=-m^2_K) = 0$ which completely fixes the counterterms at one-loop yielding a finite expression for $\Sigma_K(p^2)$ independent of any regularization parameter. The diagrams contributing to the self-energy are the following\newpage
\phantom{h}
\vspace{-1.7cm}
\begin{fmffile}{ad}
\begin{align}
\begin{gathered}
\begin{fmfgraph*}(30,30)
   \fmfleft{i}
   \fmfright{o}
   \fmf{dashes, label=$K$}{i,v}
   \fmf{dashes}{v,v}
   \fmf{dashes, label=$K$}{v,o}
   \fmfv{l=$K$,l.a=30,l.d=.3w}{i}
   \fmfv{l=$K$,l.a=150,l.d=.3w}{o}
   \fmfv{l=$\mathcal{O}(\lambda_k)$,l.a=-90,l.d=.4w}{v}
\end{fmfgraph*} \qquad 
\begin{fmfgraph*}(30,30)
   \fmfleft{i}
   \fmfright{o}
   \fmf{dashes, label=$K$}{i,v}
   \fmf{plain}{v,v}
   \fmf{dashes, label=$K$}{v,o}
   \fmfv{l=$\phi$,l.a=30,l.d=.3w}{i}
   \fmfv{l=$\phi$,l.a=150,l.d=.3w}{o}
	\fmfv{l=$\mathcal{O}(h)$,l.a=-90,l.d=.4w}{v}
\end{fmfgraph*} \qquad 
\begin{fmfgraph*}(30,30)
        \fmfleft{i}
        \fmfright{o}
        \fmf{dashes, label=$K$}{i,v1}
	\fmf{phantom,tension=.3}{i,v1}
        \fmf{phantom,tension=.3}{v2,o}
        \fmf{plain,left,tension=0.4}{v1,v2,v1}
        \fmf{dashes,label=$K$}{v2,o}
	\fmfv{l=$\phi$,l.a=52,l.d=.33w}{v1}
	\fmfv{l=$\phi$,l.a=-123,l.d=.33w}{v2}
	\fmfv{l=$\mathcal{O}(g^2)$, l.a=-55,l.d=.5w}{i}
	\end{fmfgraph*} \qquad
\begin{fmfgraph*}(30,30)
        \fmfleft{i}
        \fmfright{o}
        \fmf{dashes, label=$K$}{i,v1}
	\fmf{phantom,tension=.3}{i,v1}
        \fmf{phantom,tension=.3}{v2,o}
        \fmf{dashes,left,tension=0.4}{v1,v2,v1}
        \fmf{dashes,label=$K$}{v2,o}
	\fmfv{l=$K$,l.a=52,l.d=.33w}{v1}
	\fmfv{l=$K$,l.a=-123,l.d=.33w}{v2}
	\fmfv{l=$\mathcal{O}(g^2_k)$, l.a=-55,l.d=.5w}{i}
	\end{fmfgraph*}
\end{gathered}. 
\label{eq:diag}
\end{align}
\end{fmffile}
\noindent The first two diagrams in \eqref{eq:diag} do not have any $p^2$ dependence and are fully included in the mass counterterm. The remaining two give contributions at order $g^2$ and $g^2_k$. The associated spectral function can be calculated via analytical continuation

\be
\rho(s) =\frac{1}{\pi} \text{Im}\left(\frac{1}{-s + m^2_K - \Sigma_K(-s -i\epsilon)}\right), \qquad s=\omega^2 - {\bf p}^2.
\label{eq:ima}
\ee

\begin{figure}
\centering
\includegraphics[width=.5\textwidth]{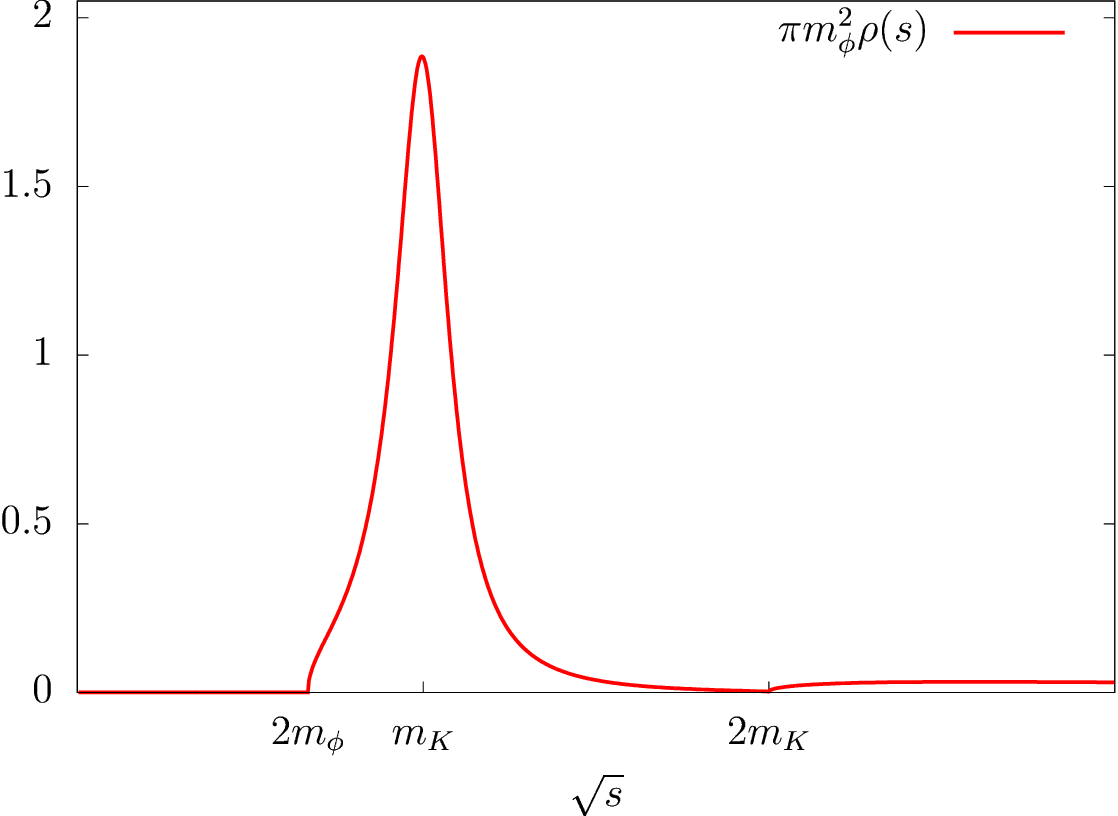}
\caption{Spectral function as defined in \protect\eqref{eq:ima} for ${\bf p}=0$, $g=15$, $g_k=80$, $m_k/m_\phi=3$.}
\label{}
\end{figure}

%\section*{Acknowledgements}

\end{document}